\documentclass[preprint2]{emulateapj-rtx4}
\usepackage{graphicx}
\usepackage{natbib}
\begin{document} 

\title{Super-star clusters versus OB associations} 

\author{Carsten Weidner, Ian~A.~Bonnell}
\affil{Scottish Universities Physics Alliance (SUPA),
School of Physics and Astronomy, University of
  St. Andrews, North Haugh, St. Andrews, Fife KY16 9SS, UK}
\email{cw60,iab1@st-andrews.ac.uk}

\and 

\author{Hans Zinnecker}
\affil{Astrophysikalisches Institut Potsdam, An der
  Sternwarte 16, D-14482 Potsdam, Germany}
\email{hzinnecker@aip.de}

\begin{abstract}
Super Star Clusters ($M_\mathrm{ecl}$ $>$ 10$^5$ $M_\odot$) are the largest stellar nurseries in our local
Universe, containing hundreds of thousands to millions of young stars
within a few light years. Many of these systems are found in
external galaxies, especially in pairs of interacting galaxies, and in
some dwarf galaxies, but relatively few in disk galaxies like
our own Milky Way. We show that a possible explanation for this
difference is the presence of shear in normal spiral galaxies which
impedes the formation of the very large and dense super star clusters
but prefers the formation of loose OB associations possibly with a less
massive cluster at the center. In contrast, in interacting galaxies
and in dwarf galaxies, regions can collapse without having a
large-scale sense of rotation. This lack of rotational support allows
the giant clouds of gas and stars to concentrate into a single, dense
and gravitationally bound system.
\end{abstract}


\section{Introduction}
\label{sec:intro}
Stars generally form in loose groups or embedded clusters in Giant
Molecular Clouds (GMC), each cluster containing a dozen to many
million of stars \citep{T78,ZMW93,LL03,Kr04b,AMG07}. These embedded clusters
need not to be bound or radially well-defined stellar ensembles and
the vast majority of them ($\sim$ 90\%) will disperse within
about 10 Myr \citep{T78,LL03}. Only a small fraction of these clusters
will hatch from the clouds as bound clusters, and after a considerable
loss in stellar numbers
due to gas expulsion \citep{Go97,GB06,WKNS06,BK08}. While these
embedded clusters vary hugely in their number of stars, their physical
sizes are all found to be rather similar with
R $\lesssim\,1$~pc~\citep{TPN98,GMP05,TLK05,RJS06,SHG07}. As the bulk of
the galactic field star populations are probably made from dissolving
embedded clusters \citep{Kr95c,LL95,LL03,AM01,AMG07}, studying the
formation and evolution of the systems is very important.

When comparing the star cluster populations of the Milky Way (MW) and
the Large Magellanic Cloud (LMC), 
we see that although the stellar populations are similar, the LMC has
much more massive clusters than
have been found in the MW. The stellar initial
mass function (IMF) do not differ beyond the expected statistical
variation \citep{Mass03,La02,WGH02,Kr02}. Furthermore, despite the
lower metallicity of the LMC and other dwarf galaxies, the mass of the
most massive stars does not seem to be any different
\citep{WK04,Fi05,OC05,Ko06,WKB09}. Surprisingly, it is in the
much smaller LMC that we find the largest stellar cluster in the local
universe, 30 Doradus. At the center of the roughly 200 pc wide
30 Doradus H{\sc ii} region lies the ``star-burst cluster'' or
super-star cluster (SSC) NGC2070/R136. It contains about half a million
stars with an estimated total mass of $\sim$~2$\cdot 10^5~
M_\odot$, extrapolated from the $\approx$~50000 $M_\odot$ in stars
from the limited range of 2.1 and 25 $M_\odot$ within an area of
about 150 pc$^2$ \citep{SMB99,AZM09}. On larger scales,
\citet{BTT08} derive a dynamical mass of 4.5$\cdot10^5~M_\odot$.
It therefore qualifies as a possible globular cluster precursor. The 
most-massive star-forming regions known in the MW have only 
$\approx$ 50000 $M_\odot$ \citep[Westerlund  1,][]{Bw08} to $\approx$
75000 $M_\odot$ \citep[Arches and Cygnus OB2,][]
{FNG02,MHP08,WSD07,NMH08,WKB09}. Though, as no strict  
definition of SSC's exist, the Arches cluster close to the Galactic
Center and NGC 3603 \citep[$\approx$ 1.3$\cdot10^4~M_\odot$,][]{HEM07}
are sometimes called SSCs. But the most massive clusters in the MW
appear to have masses 5 to 10 times smaller than 30 Doradus. It is
possible that comparable mass clusters do exist in the MW but are
obscured.

\begin{figure}
\epsscale{1.0}
\plotone{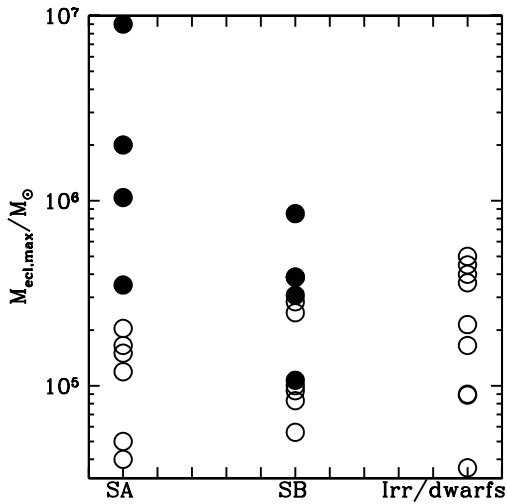}
\vspace*{-2.0cm}
\caption{Mass of the most-massive young ($<$ 50 Myr) star cluster in
  nearby spiral, barred spiral and irregular (dwarf) galaxies (see
  Tab.~\ref{tab:SFR}). Filled symbols indicate mergers and galaxies
  with signs of recent interaction.}
\label{fig:typ_vs_Meclmax}
\end{figure}

Massive or super-star clusters appear more frequent in interacting
galaxies and based on our knowledge of the local group, in dwarfs
galaxies. For example, in the interacting  Antennae
galaxies (NGC 4038/4039) \citet{ZF99} found a large number of
very massive ($\gtrsim$ 10$^6$ $M_\odot$) SSCs. In
Tab.~\ref{tab:SFR} and Fig.~\ref{fig:typ_vs_Meclmax} are shown the
most-massive young ($<$~50~Myr) star cluster for a sample of local
galaxies together with the star-formation rate (SFR) and the type of
the galaxy. For the whole sample 
there seems to be no trend with galaxy type and most-massive star
cluster but when restricted only to non-interacting galaxies (open
symbols) there is a clear trend with galaxy type. Irregular and dwarf
galaxies, which are systems with less 
shear, tend to have more massive star clusters than non-interacting
spiral galaxies. But due to the small number of objects and large
errors in the cluster masses, the statistical 
significance of the effect is low. 

This difference is even more striking when one considers that dwarf
galaxies have SFRs 5 to 10 times  
lower than do spirals, as can be seen by the mean values of the SFR
and the maximum cluster mass for different 
types of galaxies in Tab.~\ref{tab:SFR}. A naive speculation would be
that such galaxies should only produce  
lower-mass clusters, when in fact dwarfs appear to produce more
massive clusters than (non-interacting)  
spirals.

\begin{table}
{
\footnotesize
\caption{\label{tab:SFR} Most-massive young ($<$ 50 Myr) star clusters 
  in galaxies of different types. 'SA' indicates spiral
  galaxies, 'SB' barred spirals and 'Irr' irregular and dwarfs galaxies.}
\begin{tabular}{ccccc}
Name&$M_\mathrm{ecl,max}$&SFR&Type&Ref.\\
& [$10^4~M_\odot$]&[$M_\odot$ yr$^{-1}$]&&\\
\hline
NGC 45  &   3.6&0.0254&Irr&(1)\\
NGC 247 &   8.9&0.0364&Irr&(1)\\
NGC 1156&   9.0&0.186&Irr&(1)\\
NGC 1569&  40.0&0.037&Irr&(1)\\
NGC 1705&  50.0&0.03&Irr&(1)\\
NGC 5585&  21.4&0.0341&Irr&(1)\\
NGC 7793&  16.5&0.141&Irr&(1)\\
IC 1613 &   0.3&0.00045&Irr&(2)\\
LMC     &  45.0&0.2&Irr&(3)\\
SMC     &  36.0&0.15&Irr&(4)\\
mean & 23.0&0.08&&\\
\hline
NGC 300 &   5.0&0.0774&SA&(1)\\
NGC 628 &  20.4&0.993&SA&(1)\\
NGC 3184&  16.5&0.395&SA&(1)\\
NGC 5055&  11.9&0.12&SA&(1)\\
M 31    &  15.0&0.34&SA&(5)\\
M 33    &   4.0&0.21&SA&(6)\\
mean & 12.0&0.35&&\\
\hline
NGC 1313&  24.8&0.426&SB&(1)\\
NGC 2403&   9.4&0.338&SB&(1)\\
NGC 2835&   8.3&0.0924&SB&(1)\\
NGC 3521&   5.6&2.19&SB&(1)\\
NGC 7424&  28.4&0.173&SB&(1)\\
MW (Arches)& 7.7& 1.07&SB&(7)\\
mean & 14.0&0.71&&\\
\hline
NGC 2997**& 104.0&5.0&SA&(1)\\
NGC 4038/4039*& 200.0&8.3&SA&(8)\\
NGC 5194*&  35.0&4.75&SA&(1)\\
Arp 220*& 900.0&240&SA&(9)\\
NGC 3621*&  38.6&0.881&SB&(1)\\
NGC 4258*&  30.8&0.69&SB&(1)\\
NGC 5236*&  38.6&2.30&SB&(1)\\
NGC 6744*&  10.7&0.434&SB&(1)\\
NGC 6946**&  85.1&2.54&SB&(1)\\
mean*** & 68.0&3.1&&\\
\hline
\end{tabular}

* Interacting or merging galaxy. ** Only indirect evidence for a recent interaction in NGC 2997 \citep{HPW09} and NGC 6946 \citep{BOF08}. *** Arp 220 is not included in the mean values for interacting galaxies.\\
References: 1: \citet{WKL04,Lar02,Larsen2009}; 2: \citet{WHC00,Lar02};  3: \citet{AZM09,HZ09}; 4: \citet{HZ04,SSN08}; 5: \citet{CHM09,TB10}; 6: \citet{TWB02,GCG10}; 7: \citet{WKB09,RW10}; 8: \citet{ZF99,WCS10}; 9: \citet{WHL06}; 
}
\end{table}

One main difference between dwarf galaxies, disk galaxies and
interaction regions of galaxies is the amount of shear acting on
GMCs in them. While disk galaxies are rotationally supported systems,
resulting in relatively large shear forces on GMCs, dwarf galaxies
show far less amounts of shear. Interacting galaxies with tidally
driven structures are also likely to have low shear in the tidal arms and
regions of interaction where SSCs are formed (see also
\S~\ref{sec:mod} on more details about shear in disk and dwarf galaxies).

The presence of retrograde rotating GMCs \citep{B93} in spiral
galaxies is sometimes seen as evidence that the formation and
evolution of GMCs is not or only weakly connected to the
shear/rotation of spiral arms. In fact \citet{D08} has shown that
spiral shock models of 
GMC formation produce both prograde and retrograde rotating GMCs due
to the random nature of the coagulation process. Their internal shear
is still directly related to that of the  
galaxy.

In this paper we will examine the influence of different levels of
shear on the fragmentation properties of GMCs with masses of 10$^6$
$M_\odot$. The model is described in \S~\ref{sec:mod}, while
\S~\ref{sec:res} shows the results which are discussed in
\S~\ref{sec:diss}.

\section{The Model}
\label{sec:mod}

The simulations of the gravitational collapse of GMCs under different
initial conditions were carried out using a three dimensional (3D) smooth
particle hydrodynamics (SPH) code, a Lagrangian hydrodynamics
formalism \citep{Mo92}. All clouds started from cold initial
conditions in terms of both their thermal and turbulent energies being
significantly subvirial and therefore collapse rapidly due to their
self-gravity. 
A barytropic equation-of-state \citep{L05} is used with $\gamma$~=~0.75 for
densities less than $5.5 \times 10^{-19}$ g cm$^{-3}$ and $\gamma$~=~1.0
for densities above. The initial temperature is $\sim$ 50 K and the
minimum temperature reached is 7.5 K. 
The turbulent energies are $\approx
0.06$ that of the magnitude of the 
gravitational energies. The details of the initial conditions are
summarized in Table~\ref{tab:initial}. In each case the gas is
represented by 10$^6$ SPH-particles for the $10^6$ $M_{\odot}$
clouds. To model the star formation, sink particles \citep{BBP95} are
used which can grow through accretion of infalling gas (SPH particles)
and interact gravitationally with the rest of the simulation. The
radius of the sink particles is $r_\mathrm{sink}$~=~0.05~pc in all
four cases. Each sink usually starts with about 50 to 100~$M_\odot$
and can accrete up to $\sim$5000~$M_\odot$ over the length of the
simulation. The sinks are therefore not to be seen as a single stars
but rather as a small sub-clusters of stars.

\begin{table*}
\centering
\caption{\label{tab:initial} Initial conditions of the four models
  shown here. $M_\mathrm{GMC}$ is the initial mass of the GMC,
  $R_\mathrm{GMC}$ the initial radius of the GMC and $\Omega$ the
  initial angular rotational velocity. $\alpha$ and $\beta$ describe
  the following energy evolution parameters. $\alpha$ =
  $E_\mathrm{thermal}/E_\mathrm{gravitational}$, $\beta_\mathrm{turb}$
= $E_\mathrm{turbulence}/E_\mathrm{gravitational}$ and
$\beta_\mathrm{rot}$ =
$E_\mathrm{rotational}/E_\mathrm{gravitational}$. Also shown are the
mass in sinks, $M_\mathrm{4.3 Myr}$, and number of sinks,
$N_\mathrm{4.3 Myr}$, after 4.3 Myr.}
\begin{tabular}{ccccccccc}
No.&$M_\mathrm{GMC}$&$R_\mathrm{GMC}$&$\Omega$&$\alpha$&$\beta_\mathrm{turb}$&
  $\beta_\mathrm{rot}$&$M_\mathrm{4.3 Myr}$&$N_\mathrm{4.3 Myr}$\\
&$M_\odot$&pc&rad~s$^{-1}$&&&&$M_\odot$&\\
\hline
1&1$\times10^6$&50&0&4$\times10^{-3}$&0.059&0.0&4.4$\times 10^5$&1451\\
2&1$\times10^6$&50&2$\times10^{-15}$&4$\times10^{-3}$&
0.059&0.032&3.8$\times 10^5$&1326\\ 
3&1$\times10^6$&50&5$\times10^{-15}$&4$\times10^{-3}$&
0.059&0.158&2.6$\times 10^5$&1100\\ 
4&1$\times10^6$&50&1$\times10^{-14}$&4$\times10^{-3}$&
0.059&0.574&1.8$\times 10^5$&736\\ 
\end{tabular}
\end{table*}

The calculations are evolved for about one free-fall time of the
GMC. The free-fall time, $t_\mathrm{ff}$, is the time-scale an object
needs to collapse from its current radius into a single point under
the influence of gravity alone and can be given as \citep{BT87}:
\begin{equation}
\label{eq:tff}
t_\mathrm{ff} = \frac{1}{4} \left({\frac{3\pi}{2G\rho_\mathrm{GMC}}}\right)^{1/2},
\end{equation}
with $G$ being Newton's gravitational constant and $\rho_\mathrm{GMC}$
the mass density of the GMC. With a radius of $R_\mathrm{GMC}$~=~50~pc
and a mass $M_\mathrm{GMC}$~=~10$^6$ $M_\odot$ the free-fall time is
$t_\mathrm{ff}$~=~5.9~Myr in all cases considered here.

The temporal evolutions of four model-GMCs are studied, each with the
same total gas mass ($M_\mathrm{GMC}$ = 10$^6$ $M_\odot$) and radius
($R_\mathrm{GMC}$~=~50~pc) but with four different levels of
shear. The first model has no shear, while the remaining three have
shear levels corresponding to solid body rotation with angular
velocities of $\Omega = 2\times 10^{-15}, 5\times 10^{-15}$ and
$10^{-14}$~rad~s$^{-1}$. The non-rotating  model is taken to replicate the
shear level expected in either a non-rotating dwarf galaxy or one in
which the interaction with another galaxy produces regions of low
shear in the tidally-induced spiral arms. For example, in the LMC a
low shear value of $\Omega_\mathrm{LMC}\sim\,6\times
10^{-16}$~rad~s$^{-1}$ is inferred from the maximum of the rotational
velocity curve of its stars \citep{AN00}.

Models 2-4 represent typical conditions expected in spiral galaxies. The 
pattern speed of the Milky Way's spiral arms is estimated as
$\Omega_\mathrm{MW}\sim\,10^{-15}$~rad~s$^{-1}$ \citep{BEG03}. In addition,
spiral arms compress the gas from different galactocentric radii which
will increase the local shear rates by factors of 10 or more. For
example, the velocity gradient in the Orion~A GMC
\citep[$M_\mathrm{GMC}$ $\approx$ 2$\cdot10^5 M_\odot$,][]{B91} is
about $0.3\times 10^{-14}$~rad~s$^{-1}$ \citep{KTC77,Bally87}. For
several other GMCs in the Milky Way very similar values have been
measured, e.g~like 
Rosette, Mon R1 and W3 have 0.3 to $0.6\times 10^{-14}$~rad~s$^{-1}$
\citep{TLH85,B91}. Furthermore, values like the MW ones are found for
GMCs in the local spiral galaxy M33 \citep{REP03}. All these GMCs have
masses from several 10$^4$ to a few $10^5$ $M_\odot$, roughly similar
to the GMCs in our numerical study. 

The simulations are followed over $\sim$6~Myr, where the free-fall
time of the cloud is $t_\mathrm{ff}$~=~5.9~Myr (see
eq.~\ref{eq:tff}). Each model ran for about 2 month on the SUPA Altix
computer of the University of St Andrews.

The simulations presented here do not include any feedback from
supernovae, radiation or stellar winds. Though, the effects of
ionizing radiation \citep{DBC05} and stellar winds \citep{DB08} have
been studied before. We note that the inclusion of these sources of
feedback in the above models did not have a significant effect on the
star formation rate or efficiency.

\begin{figure*}
\epsscale{1.0}
\plotone{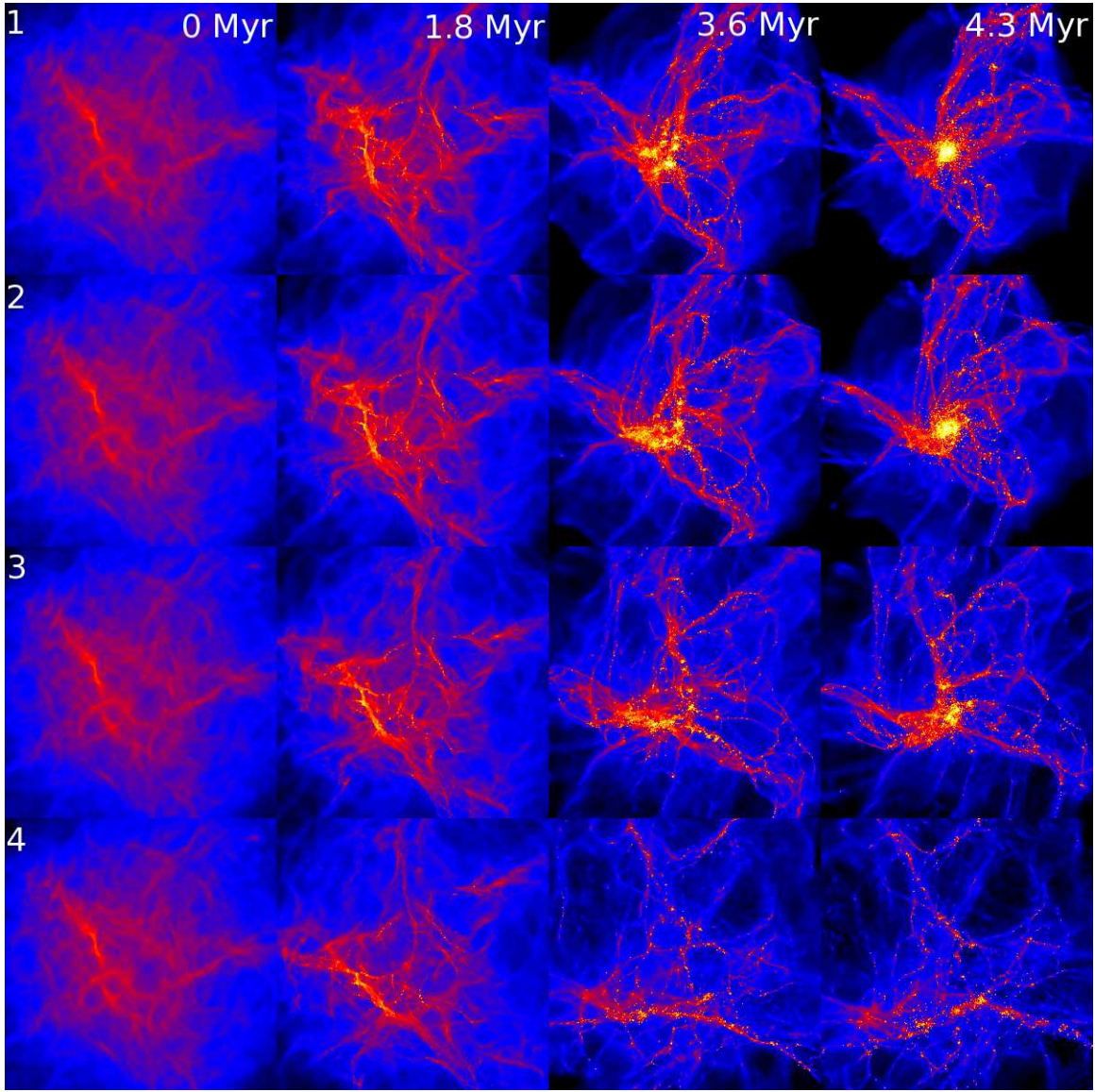}
\caption{Time-evolution series of four models. Each image
  shows a box of 60 times 60 pc. With increasing shear (model 1 to
  model 4), the amount and concentration of star-formation decreases.}
\label{fig:prettypics}
\end{figure*}

\section{Results}
\label{sec:res}

The evolution of the four models is shown in Fig.~\ref{fig:prettypics}.
The cloud evolves due to the internal turbulence, forming filamentary
structures. These structures contain local regions which are
gravitationally unstable and collapse to form sink-particles,
localized regions of star formation.  
As the collapse continues, more sink-particles are formed and these join
together in clusters which subsequently merge into one large super-cluster
containing over $10^5$ $M_\odot$ within 1 pc, and close to $4 \times
10^5$ $M_\odot$ within 10 pc.

At 4.3 Myr in Fig.~\ref{fig:prettypics} it can be seen 
that for increasing levels of shear the collapse and fragmentation
produces a more distributed population rather than the highly
concentrated super-star cluster found in the no-shear run.

All four models are evaluated at a time of about 4.3 Myr
(Fig.~\ref{fig:prettypics}) after the beginning of the calculations. As
the first sinks formed slightly after 1~Myr, some of the massive stars
will have reached an age of 3~Myr at which point they would explode
as a supernova \citep{MM03}. We halt the calculations at this point as we
do not include feedback in these models.

\begin{figure}
\epsscale{1.0}
\plotone{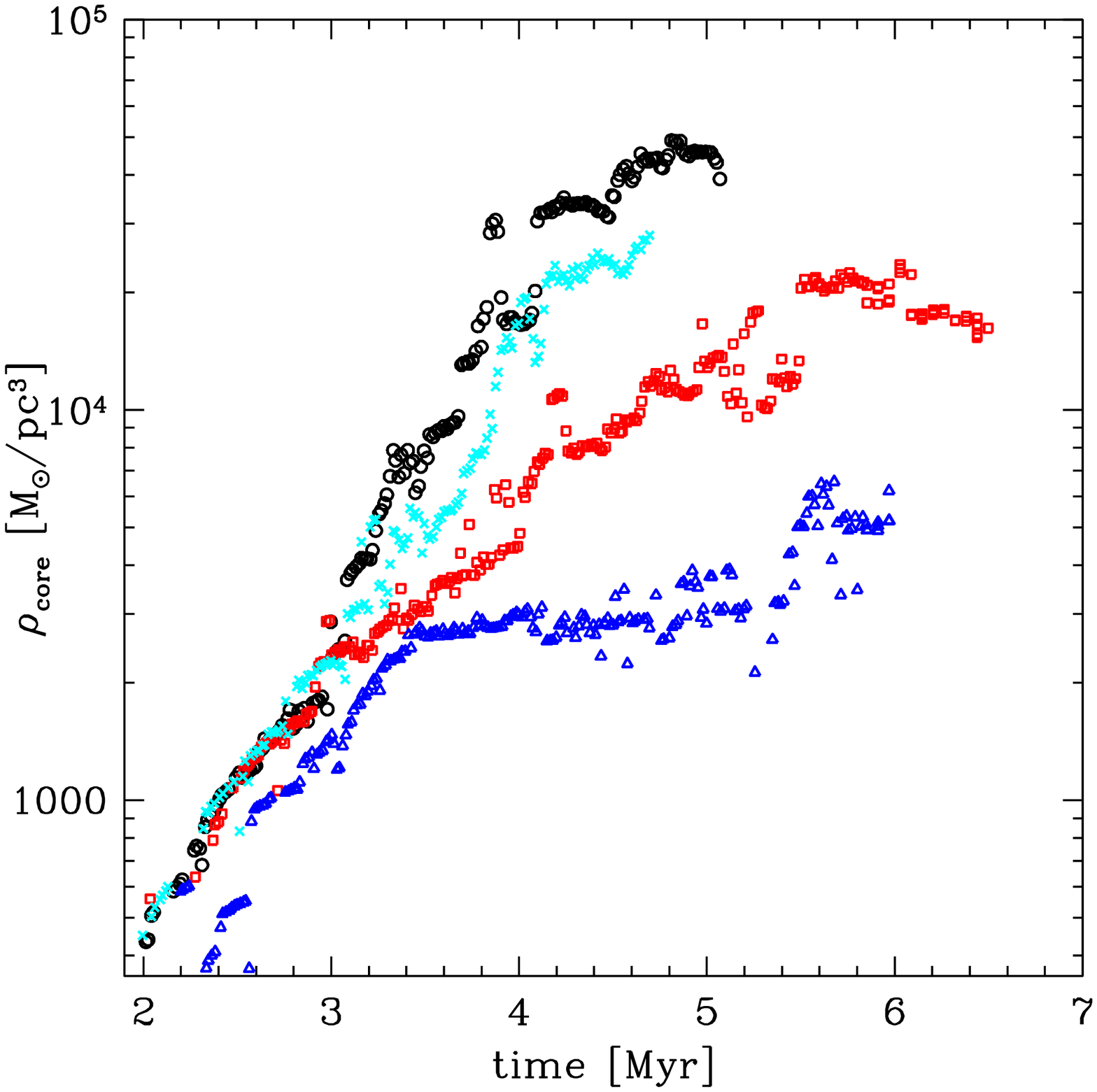}
\vspace*{-1.5cm}
\caption{Temporal evolution of mass density in a core with
  $r_\mathrm{core}$ = 1.0 pc. The black dots mark model 1, the turquoise
  crosses model 2, the red boxes model 3 and the blue triangles model
  4. After about 3 Myr of very similar evolution of all four models
the calculation without shear collapses more quickly and to higher
densities. It reaches densities which are about 3 to 15 times larger
than those in the runs with shear.}
\label{fig:noshear3}
\end{figure}

As can be seen in Fig.~\ref{fig:prettypics}, there is a strong
dependence of the amount and central concentration of stars on the
strength of the applied shear. At lower levels of shear, more sinks
are formed and they are much more  highly concentrated towards the center
of the cloud. Figure~\ref{fig:noshear3} shows the temporal evolution of
the central cluster core in each model. The mass density is calculated
in a sphere of $r_\mathrm{core}$ = 1.0 pc around the center of mass of all
sinks.  The models diverge after 3~Myr as rotational support in the shear
models halt the central collapse and core formation that occurs in the
absence of shear. The no-shear model reaches average central densities
in excess of $6\times 10^4 M_{\odot}$ pc$^{-3}$, up to 15 times larger
than the runs with shear.

\begin{figure}
\epsscale{1.0}
\plotone{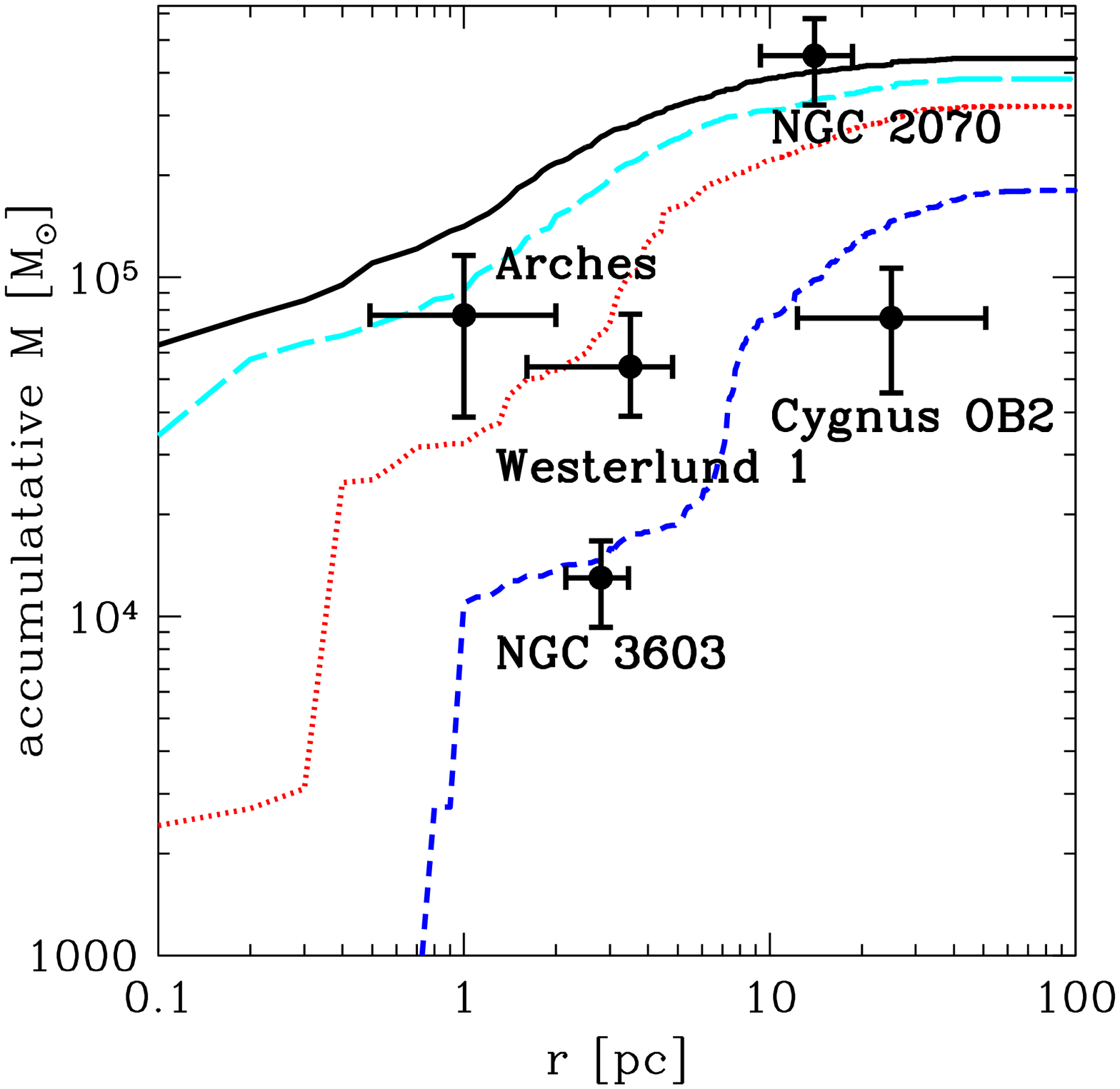}
\vspace*{-1.5cm}
\caption{The radial dependence of the accumulated mass for all four
  simulations after 4.3 Myr. The solid black line is 
  for model 1 (no shear), the long-dashed turquoise line for model 2
  (some shear), the dotted red line for model 3 (intermediate shear)
  and the dashed blue line for model 4 (high shear).
  Also shown  are several star clusters in the MW and NGC 2070 in
    the LMC. The masses are from \citet{WKB09}.} 
\label{fig:accul}
\end{figure}

\begin{figure}
\epsscale{1.0}
\plotone{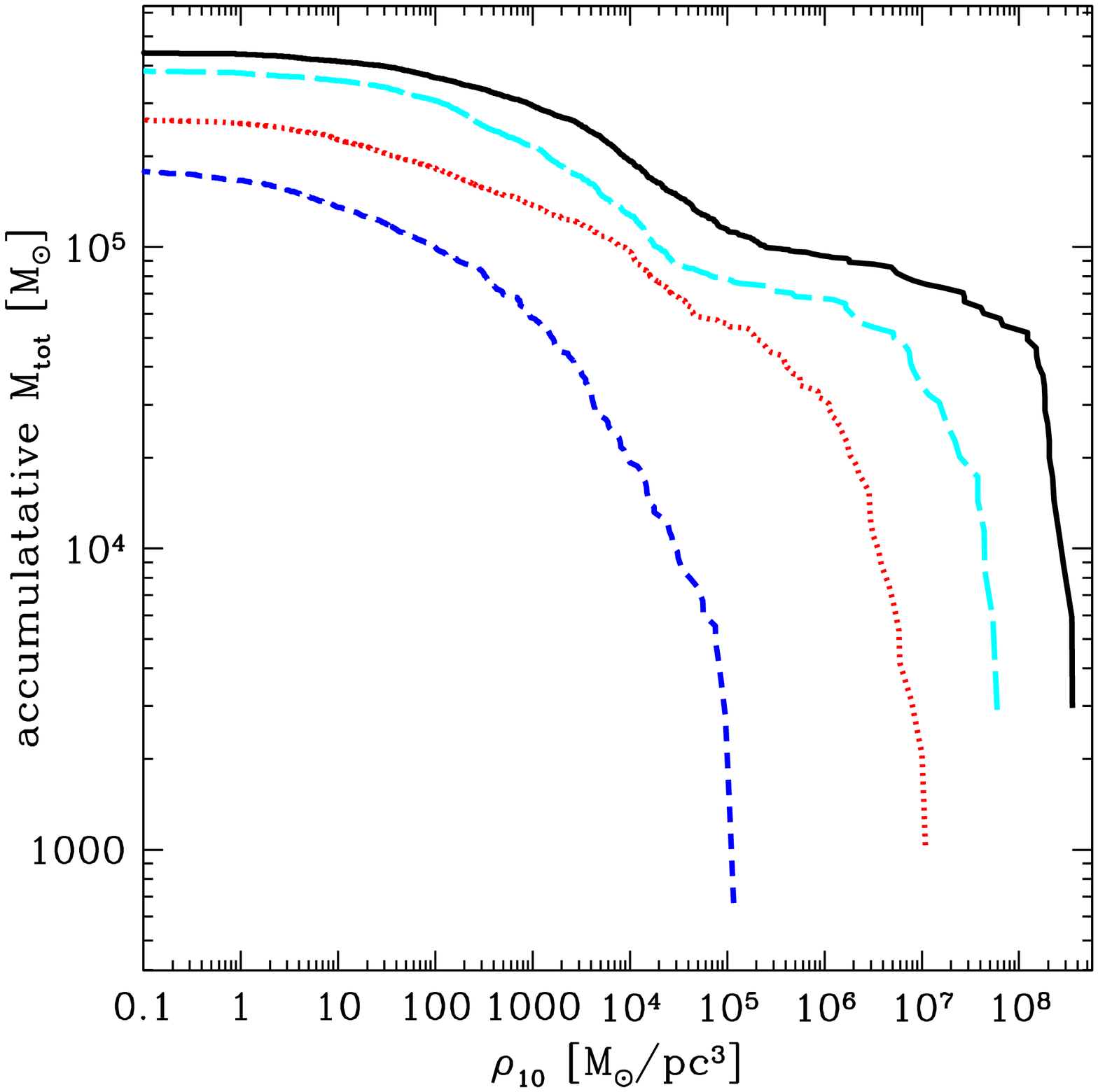}
\vspace*{-1.5cm}
\caption{The dependence of the accumulated mass on the density
  derived from the ten nearest neighbors of each sink for all four
  simulations after 4.3 Myr. The solid black line is for model 1 (no
  shear), the long-dashed turquoise line for model 2 
  (some shear), the dotted red line for model 3 (intermediate shear)
  and the dashed blue line for model 4 (high shear). Clearly, the 
SPH-calculation without shear accumulates the most mass and this mass
is very concentrated compared to the other three cases.} 
\label{fig:accul2}
\end{figure}

The difference between the models with and without shear are more evident 
when considering the mass distribution of the resultant clusters and the
distribution of local stellar densities (Figs.~\ref{fig:accul} and
\ref{fig:accul2}).
In the no shear runs, the cluster contains nearly $10^5 M_\odot$ inside
$0.1$ pc whereas the shear runs have a much more distributed
population without any significant central condensations. The
total stellar masses in all cases, and hence the star formation 
efficiencies are not drastically dissimilar  with star-formation
efficiencies (SFE = $M_\mathrm{sinks}/M_\mathrm{GMC}$) 
of 44\%, 38\%, 26\% and 18\% for model 1 to 4, respectively. It
is simply the distribution of the resultant stellar populations which
are very different. Also shown in Fig.~\ref{fig:accul} are five
star clusters and OB associations in the MW and the LMC. Especially,
the MW objects in the disk (NGC 3603 and Cygnus OB2) seem to follow
the high shear calculation quite well while NGC 2070 in the LMC fits
the low shear one. The two other MW objects, Arches and Westerlund 1,
are found in between the low and high shear run. Interestingly,
neither of these objects are in the Galactic disk but Arches is very
close to the Galactic center and Westerlund 1 is at the outer edge of
the Galactic bar, both tidally very different regions compared to the
disk.

A stronger indication is the cumulative mass distribution as
a function of local stellar density measured as the density of the ten
nearest-neighbors of each sink shown in Figure~\ref{fig:accul2}.
From this we see that approximately $10^5$ $M_{\odot}$ is in stars
which have local densities from $10^6$ to $10^8$ $M_{\odot}$ pc$^{-3}$
in the no shear run whereas in the highest shear run, less than $10^4$
$M_{\odot}$ is in stars which have local 
densities approaching $10^5$ $M_{\odot}$ pc$^{-3}$. Clearly, the 
SPH-calculation without shear accumulates the most mass and this mass
is very concentrated compared to the other three cases.

For the analysis of the simulations, each time step the center of
mass is derived by searching for the highest mass density not of the
individual sink but of the ten nearest neighbors. As can be seen in
the right panel of Fig.~\ref{fig:accul2} the local mass density
reaches rather high values ($>$ 10$^8$~$M_\odot$ pc$^{-3}$), which is
to be expected when, e.g., twenty sinks of masses $\sim\,400\,M_\odot$
are inside 0.025 pc. Even in the relatively small
\citep[$\sim\,2000\,M_\odot$,][]{HH98} Orion Nebula Cluster (ONC),
roughly comparable stellar densities are found. The projected radius
of the 4 Trapezium stars in the ONC is about 10 arc-seconds, which
translates to 0.02 to 0.025 pc for assumed distances to the ONC of 400
to 500 pc \citep{MGH08}. The combined mass of these four systems (all
four stars are binaries or higher-order multiples) is
$\sim\,100\,M_\odot$ \citep{PBH99,KWB08}. Therefore, the stellar
density is of the order of 10$^6$~$M_\odot$~pc$^{-3}$. A 
second point is that below 0.05 pc the gravity in the SPH calculation
is smoothed and hence 2-body interactions do not occur and systems are
stable where otherwise they need not to be. The sinks can cluster well
within 0.05 pc and not feel each other directly and hence be
stable. Therefore, while the local mass density might be overestimated
at times in the calculations, the main point still holds that large
amounts of mass are concentrated at high stellar densities in the
no-shear run.

\section{Discussion}
\label{sec:diss}
Our simulations show the formation of super-star clusters can depend
strongly on the shear content in the pre-collapse giant molecular cloud. 
Clouds which are only slowly rotating or
not rotating at all, as to be expected in dwarf galaxies like the LMC
and in interaction regions of colliding galaxies, can collapse
monolithically into a single massive star cluster. This is in
agreement with the results by \citet{EL08} who predict the precursors
of globular clusters to form only in galaxies not stabilized by
rotation.

In contrast, in our calculations clouds in disk galaxies are more
prone to fragmentation 
and form a system of smaller clusters or structures that could
evolve to become OB associations or relatively large clusters
susceptible to the tidal field of the galaxy  
with an extended halo of stars. Interestingly, in HST observations of
massive extragalactic clusters 
\citet{MA01} finds a similar result. Four galaxies (NGC 2403, NGC
1569, NGC 1705, LMC) of that study are  
in common with Tab.~\ref{tab:SFR}. For these, the brightest, most
compact clusters are all in dwarfs whereas  
the brightest objects in the normal spiral NGC 2403 are more extended,
classified as so-called scaled OB  
associations.

Although the results of the numerical
calculations do not exclude the possibility of SSCs formation in disk
galaxies, but they show that the presence of shear in disk galaxies
acts to impede the formation of very massive clusters. It is
therefore unlikely that the massive clusters which were the
progenitors of the present day globular clusters have formed in the
disk of galaxies. They might have formed either during an initial
monolithic collapse which formed the bulge of the galaxy, in a major
merging event, or they have been accreted from dwarf galaxies
\citep{ZKD88}.

The differences between the runs with and without shear cannot be
solely attributed to a delay in the evolution. The rotational support
in the high shear runs is not negligible and would remain important
over large timescales. Furthermore, supernova explosions would be
expected to occur in the clusters, halting any further star-formation
and potentially unbinding the cluster. But further studies including
the different feedback mechanisms are necessary to verify this
assumption.

\begin{figure}
\epsscale{0.9}
\plotone{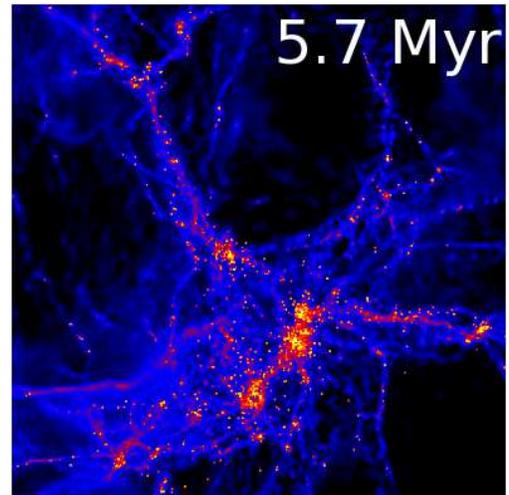}
\caption{The high shear GMC run (model 4) after 0.968 $t_\mathrm{ff}$
  (5.7 Myr). The image shows a box of 100 times 100 pc. Note the large
shell-like gas structure in the upper half of the image which has
associated star-formation along its rim. Even without any feedback in
the simulation this feature looks like a supernova blown wind-bubble
which apparently drives star-formation.}
\label{fig:prettypic2}
\end{figure}

One interesting feature of the high shear run is shown in
Fig.~\ref{fig:prettypic2}. While no feedback (like
supernovae or stellar winds) is included in the calculations, the
high-shear run has a very distinct shell or bubble like feature in the
upper part of the figure. Additionally, several sinks are found
concentrated along the rim of this feature. Such a structure is easily
mistaken as a supernova bubble. The overall appearance of the GMC is
now like a system of small star clusters with some amount of
distributed star formation around, as seen in several OB
associations in the Milky Way.

\acknowledgments
This work was financially supported by the CONSTELLATION European
Commission Marie Curie Research Training Network (MRTN-CT-2006-035890).

\bibliography{mybiblio}

\end{document}